\documentclass[fleqn,usenatbib]{mnras}

\usepackage{newtxtext,newtxmath}

\usepackage[T1]{fontenc}
\usepackage{ae,aecompl}

\usepackage{graphicx}	
\usepackage{amsmath}	
\usepackage{amssymb}	

\title[GW170817 - expansion velocity lower limit]{Cooling off with a kilonova - Lower Limit on the Expansion Velocity of GW170817}

\author[I. Linial \& R. Sari]{
Itai Linial\thanks{E-mail: itai.linial@mail.huji.ac.il}
and Re'em Sari
\\
Racah Institute of Physics, The Hebrew University, Jerusalem 91904, Israel\\
}

\date{Accepted XXX. Received YYY; in original form ZZZ}

\pubyear{2018}

\begin{document}
\label{firstpage}
\pagerange{\pageref{firstpage}--\pageref{lastpage}}
\maketitle

\begin{abstract}
GW170817 was the first detection of a binary neutron star merger via gravitational waves. The event was observed over a wide range of the electromagnetic spectrum, revealing a thermal kilonova dominating the optical signal during the first $\sim$15 days, and a non-thermal synchrotron emission that has continued to rise $\sim$200 days post-merger, dominating the radio and x-ray emission. At early times, when the kilonova is still dominant, the synchrotron emitting electrons can efficiently cool by up-scattering the kilonova photos through inverse-Compton.
Yet, the cooling frequency is not observed up to the X-ray band. This can only be explained if the source is moving at least at a mildly relativistic velocity. We find a lower limit on the source's bulk Lorentz factor of $\Gamma > 2.1$ at $9$ days. This lower limit is model independent, and relies directly on the observed quantities, providing an additional robust evidence to the relativistic motion in this event at early times.
\end{abstract}




\section{Introduction}

The first detection of a binary neutron star merger via gravitational waves, labeled GW170817 \citep{Abbott_2017_GW170817}, was accompanied by a multitude of observations across the electromagnetic spectrum, in prompt $\gamma$-ray emission \citep{Goldstein_2017_GRB}, in the UV-optical-IR band, as a kilonova/macronova, decaying during the first few days \citep{Cowperthwaite_2017,Kasliwal_2017,Drout_2017_SSS17a,Tanvir_2017,Soares-Santos_2017,Smartt_2017_KN} and a long lasting radio to X-ray afterglow that has continued to rise $\sim$200 days post-merger \citep{Hallinan_2017,Mooley_2017_radio,Alexander_2017_radio,Haggard_2017_Chandra,Troja_2017_X_ray}. 

In an effort to explain this collection of observations and the different emission components, various physical models have been considered - a very weak on-axis relativistic jet, an off-axis emission of a short-hard gamma-ray burst, interaction of the dynamical ejecta fast tail with surrounding material, a choked-jet cocoon, and a successful-jet cocoon emission \citep{Kasliwal_2017,Mooley_2017_radio,Hotokezaka_2018_Synchrotron,Gottlieb_2018_Cocoon}. Recently, \cite{Nakar_Piran_2018_Implications} have demonstrated that an off-axis relativistic jet, or an ultra-relativistic jet pointing at the observer are inconsistent with the delayed X-ray and radio emission, as well as the gradual rise of the signal, $\propto t^{0.8}$, and can be therefore ruled out.

One aspect in which the proposed models differ is the velocity of the radiating region. For instance, cocoon emission involves mildly relativistic motion of $\Gamma \sim 2-3$ \citep{Gottlieb_2017_Cocoon_emission}, whereas interaction of the merger's dynamical ejecta with surrounding material involves more modest velocities of up to $\Gamma \approx 1.4$ \citep{Hotokezaka_2018_Synchrotron}. Models that include a successful jet-cocoon ("successful structured jet") predict a Lorentz factor of up to $\Gamma \approx 10$ \citep{Kasliwal_2017,Margutti_2018_160_days,Hallinan_2017,Mooley_2017_radio}. Constraining the source velocity is therefore instrumental in discriminating between various models.

In this work we derive a lower limit on the source's velocity that is directly derived from observed quantities, without any additional free parameters. We apply this theoretical bound in the case of GW170817, and find that the emitting material must be at least mildly relativistic, with $\Gamma > 2.1$, 9 days post merger. This lower limit is consistent with recently reported limits derived from the source's centroid motion, measured between 90 and 230 days \citep{Mooley_2018_Superluminal}. Yet, our result provides an evidence of relativistic motion of the synchrotron emitting region at much earlier times.

The electromagnetic afterglow of GW170817 was identified in X-ray and radio a few days after the merger event, and has continued to rise for about $\sim$200 days \citep{Haggard_2017_Chandra,Troja_2017_X_ray,Margutti_2018_160_days,Hallinan_2017,Alexander_2017_radio}. Both bands follow the same simple power-law with no sign of any spectral evolution, implying that both originate from the same non-thermal population of relativistic electrons, emitting synchrotron radiation \citep{Sari_1998}. The fact that both radio and X-ray rise together in time, implies they are both below the cooling frequency, and above the self-absorption frequency \citep{Margutti_2018_160_days}. 

The observations in the UV-optical-IR bands during the first $\sim$15 days were dominated by a thermal kilonova emission, powered by the radioactive decay of heavy elements synthesized in the merger ejecta. At later times, as the kilonova decays, the optical band was dominated by the broadband radio to X-ray synchrotron power-law \citep{Lyman_2018_HST,Margutti_2018_160_days}.

The synchrotron-emitting electrons may cool in two different channels - losing their energy through synchrotron radiation, or alternatively by up-scattering photons via inverse-Compton. The presence of an additional photon source, i.e., the kilonova, makes inverse-Compton an important cooling mechanism at early times. The absence of a cooling frequency break in the observed spectrum sets a lower limit on the source's motion and an upper limit on the magnetic field at the synchrotron source, as we shall demonstrate.

\section{Kilonova Cooling}

\subsection{Two cooling mechanisms}
Electrons of Lorentz factor $\gamma_e \gg 1$, gyrating in the presence of magnetic field $B$ emit synchrotron radiation with power \citep[e.g.][]{Rybicki_Lightman_1986}
\begin{equation}
	P^{syn}(\gamma_e) = \frac{4}{3} \sigma_T c \gamma_e^2 \frac{B^2}{8 \pi} \,,
\end{equation}
and with a characteristic frequency in the local frame of 
\begin{equation}
	\nu^{syn}(\gamma_e) = \gamma_e^2 \frac{q_e B}{2 \pi m_e c} \,,
\end{equation}
where $\sigma_T$ is the Thomson cross-section, and $m_e$ and $q_e$ are the electron mass and charge.

Similarly, in the presence of a radiation field with energy density $u_{ph}$, the inverse-Compton power in the local frame is given by
\begin{equation}
	P^{IC}(\gamma_e) = \frac{4}{3} \sigma_T c \gamma_e^2 u_{ph} \,.
\end{equation}

Electrons that have lost most of their energy by time $T$, as measured in the local frame, have a cooling Lorentz factor of
\begin{equation}
	\gamma_c^{syn} (T) = \frac{m_e c^2}{\frac{4}{3} \sigma_T c \frac{B^2}{8 \pi} T} \,,
\end{equation}
if cooling is achieved by synchrotron emission, or
\begin{equation}
	\gamma_c^{IC} (T) = \frac{m_e c^2}{\frac{4}{3} \sigma_T c u_{ph} T} \,,
\end{equation}
if inverse-Compton is the electron cooling mechanism.

Electrons of this Lorentz factor will therefore emit synchrotron radiation at a characteristic cooling frequency
\begin{equation} \label{eq:nu_c_syn}
	\nu_c^{syn} = 18 \pi \frac{m_e c q_e}{\sigma_T^2 B^3 T^2} \,,
\end{equation}
if the electrons have cooled by synchrotron, or
\begin{equation} \label{eq:nu_c_ic}
	\nu_c^{IC} = \frac{9}{32 \pi} \frac{m_e c q_e B}{\sigma_T^2 u_{ph}^2 T^2} \,,
\end{equation}
in the case of inverse-Compton.

\subsection{Relation to observables} \label{sec:Observable_Quantities}
Consider a source, expanding towards the observer at velocity $\beta c$, corresponding to Lorentz-factor $\Gamma$. At time $t$, measured by the observer, the source has expanded to size $R = c \beta t/(1-\beta)$. This expression accounts for the travel time of photons emanating from a moving source.

Assuming that the observed bolometric luminosity of the source is $L_{bol}$, the average radiation energy density at the source local frame is given by
\begin{equation} \label{eq:u_ph_observables}
	u_{ph} = \frac{L_{bol}}{4 \pi R^2 c} \frac{1}{\Gamma^2} \approx \frac{L_{bol}}{16 \pi c^3 t^2} \frac{1}{\Gamma^6} \,,
\end{equation}
where the additional $1/\Gamma^{2}$ term appears when transforming energy density from the observer to the local frame. In the last approximation we have used $R \approx 2 \Gamma^2 c t$, valid for $\Gamma \gg 1$. In practice, the lower limit we find on $\Gamma$ is of order unity, and we use the accurate expression for $R$ rather than the approximate one. We only use this approximation for deriving simple analytical relations.

The highest measured frequency at the X-ray band, $\nu_x$ corresponds to $\nu_x/\Gamma$ at the source local frame. If no break is seen in the power-law spectrum up to this band, $\nu_c^{syn}$ and $\nu_c^{IC}$ (equations \ref{eq:nu_c_syn} and \ref{eq:nu_c_ic}, defined at the local frame) must both be greater than $\nu_x / \Gamma$. The time elapsed in the local frame ($T$ in equations \ref{eq:nu_c_syn} and \ref{eq:nu_c_ic}) is given by $T = t \beta/(\Gamma(1-\beta)) \approx 2 \Gamma t$ due to the frame transformation.

By demanding $\nu_c^{syn} > \nu_x / \Gamma$ and $\nu_c^{IC} > \nu_x / \Gamma$, one obtains an upper and lower limit on the magnetic field
\begin{equation} \label{eq:B_limits}
\frac{128 \pi}{9} \frac{\Gamma \nu_x \sigma_T^2 u_{ph}^2 t^2}{m_e c q_e} < B < \left( \frac{9\pi}{2} \frac{m_e c q_e}{\Gamma \nu_x \sigma_T^2 t^2} \right)^{1/3} \,.
\end{equation}
Equation \ref{eq:B_limits} can be rewritten with the observed luminosity $L_{bol}$
\begin{equation} \label{eq:B_limits_gamma}
\frac{1}{18 \pi} \frac{L_{bol}^2 \nu_x }{t^2} \frac{\sigma_T^2}{m_e q_e c^7} \frac{1}{\Gamma^{11}} < B < \left( \frac{9\pi}{2} \frac{m_e c q_e}{\nu_x \sigma_T^2 t^2} \right)^{1/3} \frac{1}{\Gamma^{1/3}} \,.
\end{equation}

We note that there is a minimal $\Gamma$ for which the upper limit on $B$ is greater than its lower limit

\begin{equation} \label{eq:Gamma_letters}
	\Gamma > \left( \frac{ 2^{11} \pi^{2} }{3^{8}} \right)^{1/16} \left( \frac{q^{14}}{c^{27} m_e^{10}} \right)^{1/16} \, L_{bol}^{3/16} \, \nu_x^{1/8} \, t^{-1/8} \,,
\end{equation}
where we have substituted $\sigma_T = \frac{8 \pi}{3} \left( \frac{q^2}{m_e c^2} \right)^2$. Written differently, the lower limit is given as
\begin{equation}
	\Gamma > 2.6 \, \, L_{41}^{3/16} \, \nu_{18}^{1/8} \, t_{\rm d}^{-1/8} \,,
\end{equation}
where $L_{41} = L_{bol} / \left( 10^{41} \, \rm erg \, s^{-1} \right)$, $\nu_{18} = \nu_x / \left( 10^{18} \, \rm Hz \right)$ and $t_{d} = t/\left( 1 \, \rm day \right)$.

This limit depends rather weakly on the observed quantities, and therefore provides a robust lower limit on the relativistic motion of the source.

Note that the lower limit on $\Gamma$ increases with $L_{bol}$. At a given epoch, the photon energy density at the source rapidly decreases as $\Gamma^{-6}$ for a given bolometric luminosity measured by the observer (equation \ref{eq:u_ph_observables}). This is mostly since higher $\Gamma$ corresponds to larger source size, and a hence a lower photon energy density at the source. Since the inverse-Compton cooling rate is proportional to $u_{ph}$, the absence of a cooling break in the measured spectrum implies that cooling by inverse-Compton is sufficiently slow. The efficiency of inverse-Compton cooling can therefore be drastically reduced when $\Gamma$ is sufficiently large, setting the lower limit of equation \ref{eq:Gamma_letters}.

\subsection{Kilonova as the photon source for inverse-Compton}
GW170817 was followed by a prompt $\gamma$-ray emission, a short lived thermal kilonova and a long lasting synchrotron components. At early times, up to $\sim$15 days, the luminosity is dominated by the kilonova, which provides photons for electron cooling through inverse-Compton. In this section we argue that the kilonova radiation inevitably interacts with the synchrotron source.

The spectral evolution during the first days of observations constrains the photospheric expansion velocity of the kilonova to be $0.1c$ to $0.3c$ \citep{Kasliwal_2017}. Hydrodynamic simulations of neutron star mergers demonstrate that the merger ejecta, which is the kilonova emitting region, expands quasi-spherically \citep{Hotokezaka_2013,Bauswein_2013}. The kilonova's thermal emission therefore originates from a slowly expanding spherical ejecta.

As we demonstrate for GW170817, the synchrotron source must be at least mildly relativistic with $\Gamma > 2.1$, and it is therefore further out than the slower kilonova emitting ejecta, which expands isotropically. Whether the synchrotron emission comes from a narrow jet, a wide angle cocoon or a spherically expanding shell, a scenario in which the synchrotron source evades the kilonova photons is unlikely. The synchrotron emitting electrons are irradiated by the kilonova, and cool down by inverse-Compton upscattering of these photons, making our derivation self-consistent.

\section{Results}
The first detected optical counterpart to GW170817, SSS17a was identified in the galaxy NGC 4993, located at a distance of $40 \, \rm Mpc$. The bolometric luminosity was estimated to be $\sim 10^{42}  \, \rm erg \, s^{-1}$ at $t = 0.5 \, \rm d$, and has declined in time roughly as $t^{-0.85}$ \citep{Drout_2017_SSS17a}.

Simultaneous observations of radio, optical and X-ray were available starting from $t = 9 \, \rm d$ \citep{Hallinan_2017,Tanvir_2017,Soares-Santos_2017,Cowperthwaite_2017,Kasliwal_2017,Margutti_2018_160_days}. At this epoch, no cooling-frequency break is seen up to $\nu_x = 2.4 \cdot 10^{18} \, \rm Hz$, with radio and X-ray measurements consistent with the same power-law flux density $F_{\nu} \propto \nu^{-0.6}$. The bolometric luminosity at that time is $L_{bol} \approx 6 \cdot 10^{40} \, \rm erg \, s^{-1}$. Using inequality \ref{eq:Gamma_letters} from section \ref{sec:Observable_Quantities}, we obtain
\begin{multline}
	\Gamma > 2.0 \, \left( \frac{L_{bol}}{6 \cdot 10^{40} \, \rm erg \, s^{-1}} \right)^{3/16} \left( \frac{\nu_x}{2.4 \cdot 10^{18} \, \rm Hz} \right)^{1/8} \left( \frac{t}{9 \, \rm d} \right)^{-1/8} \,.
\end{multline}

When taking the accurate expression for the source radius (see equation \ref{eq:u_ph_observables} and the afterwards discussion), we obtain a somewhat higher lower limit for this event, of $\Gamma > 2.1$.

The lack of a cooling break, combined with the availability of kilonova photons that can cool down the source's electrons via inverse-Compton, implies that the source must be moving at a relativistic velocity at this time. Since the electrons have not cooled down efficiently via synchrotron either, we find an upper limit on the magnetic field (right hand side of inequality \ref{eq:B_limits_gamma})
\begin{multline}
	B < 5.2 \cdot 10^{-3} \, \left( \frac{L_{bol}}{6 \cdot 10^{40} \, \rm erg \, s^{-1}} \right)^{-1/16} \\ \left( \frac{\nu_x}{2.4 \cdot 10^{18} \, \rm Hz} \right)^{-3/8} \left( \frac{t}{9 \, \rm d} \right)^{-5/8} \, \rm gauss \,.
\end{multline}

Note that the limit on $\Gamma$ becomes less constraining at later times. As $t$ increases, the radius is larger for a fixed expansion $\Gamma$, and additionally the kilonova's luminosity rapidly decays with time. Both these trends contribute to a smaller photon energy density at the source, $u_{ph}$, and thus lower $\Gamma$ is required to set the cooling-frequency above the X-ray band.

The limit we obtained on $\Gamma$ is in agreement with various estimates of $\Gamma$ found by different authors. \cite{Margutti_2018_160_days} use the precise measurement of the synchrotron's spectral index to constrain $\Gamma \sim 3 - 10$. Their estimate is based on \cite{Keshet_Waxman_2005}, who obtain the spectral index of particles accelerated in relativistic collisionless shocks. Models explaining the observations as cocoon emission imply that the emitting region is moving at $\Gamma \sim 2-3$ \citep{Gottlieb_2018_Cocoon}. Compactness arguments linked with the observed prompt $\gamma$-ray emission spectrum originally suggested that $\Gamma \gtrsim 2.5$ \citep{Kasliwal_2017,Gottlieb_2018_Cocoon}. Recently, VLBI observations have demonstrated superluminal centroid motion of the synchrotron source at late time, implying $\Gamma \gtrsim 4$. The VLBI observations, combined with compactness arguments set $\Gamma \gtrsim 5$ as a lower limit of the $\gamma$-ray emitting region \citep{Matsumoto_2018}. \cite{Nakar_Piran_2018_Implications} have shown that the gradual rise of the lightcurve implies an on-axis source. They estimate the Lorentz factor of the emitting region from the observed flux to be $\Gamma \approx 1.5-7$. The uncertainty in their estimate comes mostly from the unknown density of the surrounding material and the magnetic field's equipartition parameter.

Our result disfavors models in which the source is moving at sub-relativistic velocities, such as the one proposed in \cite{Hotokezaka_2018_Synchrotron}, where the fast tail of the merger's dynamical ejecta, moving at $\Gamma \sim 1.4$, is invoked as the source of the non-thermal synchrotron emission.

\section{Discussion and Conclusions}

Recently, radio observations revealed that the source is moving superluminally \citep{Mooley_2018_Superluminal}, setting $\Gamma \gtrsim 4$ as a lower bound on the source's motion, higher than the limit we obtained. The centroid motion of the source has been estimated between 75 days and 230 days post-merger. Notwithstanding, GW170817 occurred at merely $40 \, \rm Mpc$, well within the LIGO-Virgo detection horizon. The majority of future detections will be of merger events at greater distances, nearing the outskirts of LIGO's sensitivity horizon. Measuring the centroid motion of most future events will therefore prove more challenging than in the case of GW170817, due to limited angular resolution of radio observations.

On the contrary, the limit we obtain is derived from the observed fluxes across different spectral bands, and does not rely on significant displacement of the source. Our limit can be particularly useful for characterizing sources that are sufficiently bright across the different bands at early times, but are too distant to resolve or measure their centroid motion. Additionally, our approach constrains the source's motion in early times, as opposed to radio observations of the source's motion, which may only be feasible at late times, after the source has expanded or moved sufficiently.

The theoretical limit we have obtained will be useful in future events in which early time observations will show that the cooling frequency is above the X-ray band, despite the presence of a bright kilonova signal. The limit we found in this work will then constrain the source's expansion velocity at early times, instrumental in discriminating between different physical scenarios.

The kilonova that followed GW170817 has decayed in time, with its luminosity decreasing roughly as $\propto t^{-0.85}$ \citep{Drout_2017_SSS17a}. Therefore, at earlier times the photon energy density at the source was greater, increasing the cooling rate by inverse-Compton. One could have set a more stringent lower limit on $\Gamma$, if broadband radio to X-ray observations of this event were available at times earlier than $t = 9 \, \rm d$. If the cooling frequency was above the X-ray band at time $t$, the lower limit on $\Gamma$ is
\begin{equation}
	\Gamma_{min} = 3.6 \left( \frac{t}{1 \, \rm d} \right)^{-0.28} \,.
\end{equation}

In cases where the X-ray emission is detected at very early times, an inverse-Compton cooling frequency may be visible in the data. This occurs due to the relative small source size, and the enhanced kilonova brightness at these early times. The cooling break will later disappear, when the kilonova fades, and may reappear in later times, when synchrotron cooling becomes the dominant cooling mechanism.

Finally, the results obtained in this work can be applied to other types of events. One interesting example is the afterglow of long GRBs that are associated with supernovae. The supernova contributes an excess of radiation energy density at the afterglow source. If no cooling break is seen up to some frequency, one can constrain the Lorentz factor of the synchrotron emitting region of the afterglow.

\section*{Acknowledgements}
IL thanks support from the Adams fellowship. RS is supported by an ISF and an iCore grant. We thank Tsvi Piran, Kenta Hotokezaka and Sivan Ginzburg for comments and discussion.




\bibliographystyle{mnras}
\bibliography{gw170817_gamma}





\bsp	
\label{lastpage}
\end{document}